# Numerical Analysis of a Highly Sensitive SOI MRR Refractive Index Sensor with Performance Enhancement using Graphene and Gold

Tasin Intisar[1], Ahmed Shadman Alam[1], Ishtiaqul Hoque[1], Md Omar Faruque[1,*]

[1]Department of Electronics and Electrical Engineering, Islamic University of Technology, Board Bazar, Gazipur -1704, Bangladesh.

[*]Corresponding author email: omarfaruque@iut-dhaka.edu

*Abstract* - **This study proposes a simulation-based design for a Silicon-On-Insulator (SOI) ring resonator with a Figure of Merit (FOM) of 56.15 and a high sensitivity of up to 730 nm/RIU. The Finite-Difference Time-Domain (FDTD) technique was used to assess and evaluate the design quantitatively. Our design demonstrates higher sensitivity compared to many recent studies conducted on SOI-based sensors. The device structure follows a conventional ring resonator arrangement with a single waveguide, incorporating a 2D graphene layer on top of the SiO2 wafer and a gold nano-disc positioned at the center of the ring. Our findings highlight the device's susceptibility to refractive index variations, making it a desirable choice for various sensing applications. We have investigated the sensor's capabilities for sensing different concentrations of milkmilk. Graphene and gold materials enhance the device's response to light and provide comparatively higher sensitivity. The suggested design can serve as a blueprint for device fabrication, considering the practicality of implementing an SOI-based device using standard techniques for silicon processing.**

## 1. Introduction:

Micro-ring resonators (MRRs) made of silicon-on-insulator (SOI) have become susceptible instruments used in optical sensing applications [1]. To obtain wavelength-selective filtering or sensing capabilities, SOI MRRs use the resonant behavior of light in a circular waveguide constructed on an SOI substrate [2]. A silicon waveguide with a closed loop and a short radius, often in the micrometer range, makes up the SOI MRR, a form of waveguide-based optical resonator [3]. SiO$_2$ and Si have very different refractive indices; hence, this substance can function as a highly restricted optical waveguide [4]. The ring resonator's unique geometry makes it possible to confine and recirculate light waves, producing resonant modes with specific wavelengths or frequencies [2].

MRRs, both SOI and plasmonic metal insulator metal (MIM), are appropriate for a variety of sensing applications [5]–[8], including biosensing, environmental monitoring, and chemical sensing [2], [9], [10], because these resonant modes are susceptible to changes in the refractive index of the surrounding medium.

Refractive index sensors can precisely detect changes in the refractive index of unidentified materials that have recently received much attention. Waveguide designs built in all other standard planar material systems cannot match the sensitivity of sensors based on SOI [11]. The essential characteristics for assessing the performance of a Refractive Index (RI) sensor are its sensitivity and FOM [12]. FOM is defined as sensitivity divided by Full Width at Half Maximum (FWHM) [12]. FWHM plays a crucial role in establishing the quality of a sensor, as a lower FWHM allows us to observe a more distinct resonance shift. Refractive index sensors are also well suited for the development of numerous sensor categories for evaluating biosensing, including blood group detection[13]–[16], hemoglobin concentration detection [17], and cancer cell detection [18].

As a result of earlier research, additional attempts have been made to improve the performance of SOI MRRs by introducing graphene, a two-dimensional material with outstanding characteristics. High-performance optical modulators and biochemical sensors can be made from graphene, which can also impact the transmission spectrum [13]. Various wavelengths, including the visible and near-infrared, show substantial optical absorption in graphene. The graphene layer can significantly absorb some of the incident light when positioned close to the SOI MRR, improving the light-matter interaction inside the device. The signal strength can be amplified, and this higher absorption can enhance the SOI MRR's overall performance.[19]. The SOI MRR structure's incorporation of graphene improves the interaction between light and the environment. Surface plasmons and other excitations can be created on the graphene layer and then coupled with the guided modes of the MRR. This coupling improves the guided mode's evanescent field interaction with its surroundings, making it more sensitive to changes in refractive index or other analyte-specific characteristics. As a result, the SOI MRR's overall performance, sensitivity, and detecting skills can be enhanced [20].

Apart from Graphene, Gold has also been added separately to the structure to test its potential for enhancing sensing performance. Gold is recognized for its plasmonic capabilities. Due to the metal-dielectric interaction supporting the excitation of Surface Plasmon Polaritons (SPPs), a hybrid nature is manifested. Electromagnetic waves known as SPPs are produced by the interaction of light photons and the surface electrons of metals as they traverse across their surfaces [14], [21], [22]. The SPPs induce an evanescent electromagnetic field around the gold nano-disc. This field extends into the surrounding medium, into the nearby SOI ring resonator gold's near-infrared and visible wavelength regions. Ultimately, the interaction between the evanescent field of the SPPs and the Si ring resonator results in resonance coupling.

Moreover, gold's near-infrared and visible wavelength regions show negligible optical losses compared to other metals like Silver. It also has an elevated oxidation and corrosion resistance level, making it excellent for long-term device stability [23]. Additionally, gold is biocompatible. Due to this characteristic, gold-based resonators are useful for biosensing and diagnostic applications that need interaction with biological materials [24]. However, some loss might be caused by the gold nano-disc in the resonators, which creates chemical interface damping between different layers and materials used [25].

Recently, many studies on SOI MRR refractive index sensors have been undertaken. Cimenelli et al. [26] suggest SOI microcavities based on a planar design with a 120 nm/RIU sensitivity. The sensitivity for a slot ring resonator, consisting of two rings with a slot in between, was observed to be 360 nm/RIU by Raj et al. [27]. A hybrid SOI MRR with a 555 nm/RIU bulk refractive index sensitivity was studied by Kumari et al. [24]. These designs show increasing sensitivity, but there was still an opportunity for development regarding lab-on-chip device optimization.

Compact and straightforward to build, MIM waveguides have a remarkable balance between light localization and propagation loss [28][29]. MIM waveguides have received significant attention because they can surpass the diffraction limit of light [30]. Some configurations of these types of plasmonic sensors include an elliptical resonator [31], an H-shaped cavities refractive index sensor [32], and a plasmonic metal-based sensor structure that is coupled with an aperture to the concentric square disc and ring resonator [33]. Recent theoretical studies have studied a high-sensitivity plasmonic refractive index sensor for bio-chemical detection, with SPPs induced by the gold nano-disc and Si ring resonator interaction [34]. Fei et al. and Chen et al. present their works based on multiple Fano resonances in a plasmonic waveguide resonator system with a nanorod-defect [35] and MIM waveguide system coupled with a resonator structure[33], respectively. Based on Fano resonance, Zafar et al. [37], Sharma et al. [38], and F. Chen et al. [39] have undertaken more such works. Similarly, optical refractive index (RI) sensors based on a photonic crystal were also designed by Leila et al. [40] and M.R. Islam et al. [41].

Although MIM sensors have been found to possess higher sensitivity, it is noteworthy that numerous investigations of these sensors have exclusively relied on 2-D simulations, which may not comprehensively encapsulate their intricate characteristics. 2-D simulations can be misleading compared to 3-D modeling or experimental outcomes. Our study's investigation of SOI sensors has utilized 2.5D simulations incorporating vertical structures, resulting in more accurate sensing parameter values. Using 2.5D simulations of Ansys Lumerical Photonics Simulation & Design Software can yield outcomes like those obtained from 3D simulations in a shorter time, thereby mitigating certain constraints inherent to 2D simulations and providing values closer to experimental works and comparable to 3D simulation results. The utilization of 2.5D simulations for SOI sensors in this study enhances the precision of comprehending their performance attributes. Studies by Alina et al. [42] and Z. Tu et al. [43] present simulations with experimental results of sensitivity much lower than theoretically investigated plasmonic sensors.

Additionally, the paper's architecture may be directly compared to a planar sensor but with increased sensitivity. Simple geometry was used in the modeling, and it may be arranged periodically to create massive metamaterial structures with several of these resonators for THz sensing with increased sensitivity[44]–[47]. However, since this further adds design complexity, a better sensing capacity for our suggested structure would help to address this problem and give acceptable sensing accuracy with a more simplistic model.

In the present work, a comprehensive investigation of an overly sensitive hybrid SOI MRR with graphene and gold for optical sensing applications was made, where the resonant behavior and performance of the device were using the plasmonic properties of gold and graphene.

We improved the SOI MRR's design characteristics to obtain optimum sensitivity in this simulation. A regular MRR was tested, and a maximum sensitivity of 730nm/RIU with a complementary FOM of 56.15 was obtained after optimization of the structure and enhancing the performance by adding gold and graphene. The basic design had a circular ring cavity and a straight rectangular waveguide, where the fundamental mode of the source was applied to the input port. The theoretical Finite Difference Time Domain (FDTD) method was utilized to simulate the MRR and analyze the performance parameters using the software. FDTD makes it possible to calculate both the spatial and temporal features in a single run, making it an effective tool for analyzing various structures[48]. To the best of our knowledge, the significantly large sensitivity value is the highest observed for an SOI MRR, making the device most suited for on-chip refractive index nano-sensors.

## 2. Structural design and computation:

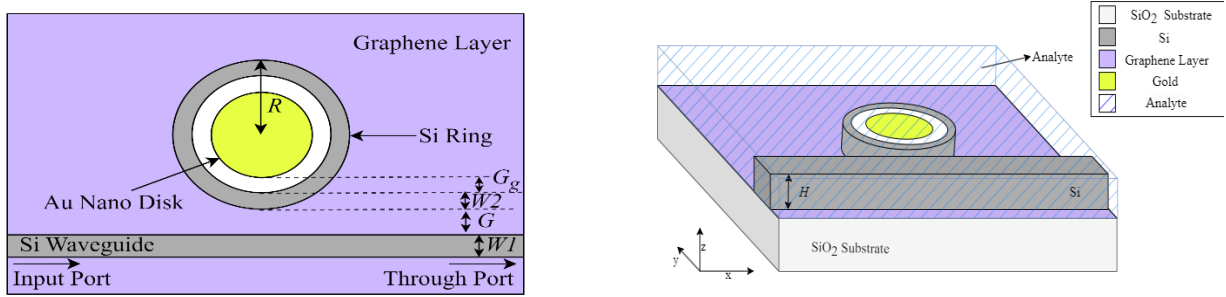

**Fig. 1.** Model of the proposed MRR in (a) 2D and (b) 3D.

Our proposed structure has been illustrated in Fig. 1 (a) in 2D top view and Fig. 1 (b) in 3D, where a micro-ring resonator and a rectangular waveguide have been coupled. Table 2 exhibits the geometrical characteristics of the SOI MRR before optimization. The MRR and the straight rectangular waveguide are silicon, whereas the substrate is $SiO_2$. $W_1$ denotes the width of the rectangular waveguide, and $W_2$ indicates the width of the ring resonator. $G$ represents the distance between the straight waveguide and the MRR, while $H$ denotes the total height. After optimization, a single graphene layer was deposited on the substrate, followed by gold (Au) incorporation within the MRR.

The fundamental operating principle of the MRR is based on whispering gallery mode (WGM) [49]. In sensors based on WGM optical resonators, light propagates as WGM resonances. These resonances result from light's total internal reflection along the curved surfaces of the resonator

[50]. The WGM pattern generates a strong evanescent field at the interface between the waveguide and the ambient medium, with the strongest region occurring at the ring's equator. This field permeates the surrounding medium and interacts with any analyte molecules or particles that may be present. The interaction alters the effective index of refraction, $n_{eff}$ of the optical WGM resonance. As a result, the cavity WGM resonance wavelength $\triangle \lambda_{WGM}$ experiences a net spectral shift. The well-established resonance condition links $\triangle \lambda_{WGM}$ and $\triangle n_{eff}$ using the relationship in equation (1) [50]–[52]:

$$\Delta\lambda_{WGM} = \frac{2\pi R}{m} \Delta n_{eff} \qquad (1)$$

where $R$ stands for the ring radius, m for the integer mode order, and $\triangle \lambda_{WGM}$ stands for the wavelength [50]. Equation (1) only asserts that the resonance condition is met when an integer number of wavelengths equals the length of the round trip around the cavity [53]. The measurement of $\triangle \lambda_{WGM}$, which $\triangle n_{eff}$ causes, makes it feasible to detect analytes. So, to proceed, we must first be able to determine the precise value for $\triangle n_{eff}$ that the WGM resonance wavelength experiences in both the air and biological claddings [50]. The transmission spectrum of the input waveguide is used to track the resonances of WGM cavities [53].

The coupling between the ring resonator and the excitation signal ($E_{i1}$) provided at the input port are related by equation (2) [13], [54], [55]:

$$\begin{pmatrix} E_{t1} \\ E_{t2} \end{pmatrix} = \begin{pmatrix} t & k \\ -k* & t* \end{pmatrix} \begin{pmatrix} E_{i1} \\ E_{i2} \end{pmatrix} \qquad (2)$$

$E_{t1}$ and $E_{t2}$ are the normalized amplitudes of complex modes. The couplers' transmission coefficient (*t*) and coupling coefficient (*k*) rely on one another based on equation (3) [54]:

$$|t^2| + |k^2| = 1 \qquad (3)$$

Equation (4) shows how waves propagate inside the ring segment of the resonator:

$$E_{i2} = E_{t2} L e^{-j\phi} \qquad (4)$$

where $\phi$ is the accumulated phase, and $L$ is the loss coefficient for the propagation around the ring. The expression for the transmission coefficient and its phase is given by equation (5):

$$t = |t| e^{-j\phi_t} \qquad (5)$$

The transmitted field expression from equations (2), (3), (4), and (5) is expressed as equation (6),

$$E_{t1} = \frac{|t| - Le^{-j(\phi-\phi_t)}}{1-|t|Le^{-j(\phi-\phi_t)}} E_{i1} e^{\phi_t} \qquad (6)$$

The transmission loss /t/ and the corresponding phase $\phi_t$ are two components of the transmission coefficient [54].

Graphene, a single-layer material, exhibits a distinctive two-dimensional structure without a band gap [56]. When a graphene monolayer is deposited on the model, it modifies the effective refractive index due to the analyte's presence. The SOI ring resonator's evanescent field can be altered by varying the external electric field while tuning graphene's refractive index. Furthermore, surface plasmons that move along the graphene surface are facilitated by the enhanced carrier dynamics of graphene. These plasmons can interact strongly with the ring resonator's evanescent field and significantly change its characteristics to improve sensitivity [57].

So, the optical properties of graphene and gold are essential for a refractive index sensor. Graphene plasmonic exhibits low Ohmic loss, which is critical for developing powerful plasmonic tunable devices. Graphene plasmonic exhibits superior characteristics to traditional metal plasmonic materials, which can be conveniently adjusted through chemical doping or gate voltage. Kubo's formula in equation (7) can compute the complex surface conductivity, a crucial electrical parameter of graphene [58], [59],

$$\sigma_{S,G}(\omega, \Gamma, \mu_c, T) = \sigma_{inter}(\omega, \Gamma, \mu_c, T) + \sigma_{intra}(\omega, \Gamma, \mu_c, T) \qquad (7)$$

where $\sigma_{inter}(\omega, \Gamma, \mu_c, T)$ and $\sigma_{intra}(\omega, \Gamma, \mu_c, T)$ represent the absorption corresponding to the inter-band electron transition and intra-band electron-photon scattering, respectively. Equations (8) and (9) can be used to compute these parameters:

$$\sigma_{inter}(\omega, \Gamma, \mu_c, T) = j\frac{q_e^2(\omega + 2j\omega\Gamma)}{\pi\hbar^2} \int_0^\infty E\left(\frac{f_d(-E)}{(\omega + j2\Gamma)^2} - \frac{f_d(E)}{4\left(\frac{E}{\hbar}\right)^2}\right) dE \qquad (8)$$

$$\sigma_{intra}(\omega, \Gamma, \mu_c, T) = \frac{-jq_e^2}{\pi\hbar^2(\omega + j2\Gamma)} \int_0^\infty E\left(\frac{\partial f_d(E)}{\partial E} - \frac{\partial f_d(-E)}{\partial E}\right) dE \qquad (9)$$

here $\tau$ is the relaxation time, which is correlated with carrier mobility, $\hbar$ is the reduced Planck constant, $\omega$ is the angular velocity, $k_B$ is the Boltzmann constant, $q_e$ is the electron charge, $T$ is the temperature, $\mu_c$ is the chemical potential of the doped graphene material, and $\Gamma$ is the scattering rate, $f_d$ is the Fermi-Dirac distribution and $E$ represents energy. In our study, the thickness of the graphene is negligible; thus, the bulk conductivity can be found using the simplified equation (10):

$$\sigma_{V,G}(\omega, \Gamma, \mu_c, T) = \frac{\sigma_{V,G}(\omega, \Gamma, \mu_c, T)}{\delta} \quad (10)$$

In the stationary regime, the complex relative bulk permittivity of graphene is given by Ohm's law and Ampere's law, and this can be calculated from equation (11) [59], [60]:

$$\varepsilon_{V,G} = 1 + j\frac{\sigma_{V,G}(\omega)}{\varepsilon_o \omega} \quad (11)$$

Our study examines a graphene material model with the following parameters: The temperature was set at 300K, the chemical potential was at 0.65 eV, and the scattering rate was 0.00051423eV.

The dielectric constants of different materials change with frequency. The Drude model may represent the frequency-dependent permittivity of gold given by equation (12) [61].

$$\varepsilon_m(\omega) = \varepsilon_\infty - \frac{\omega_p^2}{\omega(\omega + i\omega_c)} \quad (12)$$

where $\varepsilon_\infty$ is a constant that takes into consideration inter-band transitions. $\omega_p$ indicates the temperature-dependent plasma frequency. $\omega$ is the angle of frequency of incident vacuum light. $\omega_c$ denotes the temperature-dependent collision frequency. Empirical figures were obtained from Johnson et al.'s study [61].

A gold layer in the shape of a disc has been placed on top of the silicon dioxide layer and surrounded by the silicon ring to boost sensitivity further. The evanescent field emerging from the ring and the gold layer can interact strongly. The gold disc interacting with the evanescent field region induces the excitation of localized surface plasmons, which substantially enhances the electric field intensity at the surface of the gold layer. Compared to the incident amplitude, the total electric fields near-gold disc amplitude can significantly increase at resonance. Due to the field accumulation in the gap region, high sensitivity values can be obtained. With the addition of gold, the analyte's effect on the effective refractive index is boosted, thus increasing the sensitivity [53].

## 3. Results and Discussions:

*3.1 Simulation and Result Calculation Method:*

The property of the SOI MRR to sense a shift in the refractive index of dielectric is one of the key features broadly utilized. Two parameters that result from the refractive index sensing capabilities of the SOI MRR, Sensitivity (S) and FOM, are essential to using these devices as sensors. The sensitivity depends on the resonant wavelength peak shift due to the material's refractive index range. It can be calculated using equation (13) [22],

$$S = \frac{d\lambda}{dn} \quad (13)$$

On the other hand, the resonance spectrum's Sensitivity and Full Width Half Maxima (FWHM) can be used to compute FOM. To express FOM, we use the equation (14) [22],

$$FOM = \frac{S}{FWHM} \quad (14)$$

The sensitivity increases with a bigger FOM while lowering the FWHM bandwidth. For lower FWHM values, the notch of the transmission profile is smaller, which enhances detection. Therefore, high sensitivity and FOM are necessary for an optical RI sensor to be more effective.

The measurement used to determine the sensor's detection limit, or the slightest change in refractive index that the sensor can detect, is known as sensing resolution. It is defined in equation (15) [22],

$$SR = \left(\frac{dn}{d\lambda}\right)\Delta\lambda \quad (15)$$

$\Delta\lambda$ is the wavelength resolution.

To conduct our simulation of the SOI MRR, we had to maintain certain controls and apply specific boundary conditions while using the simulation software. We applied a metal layer for boundaries in the z-direction, while perfectly matched layer (PML) conditions were applied in the x and y directions. The type of PML we used was the stretched coordinate PML. In our study, we employed the auto non-uniform mesh type. The mesh size was set to 0.00025 µm, with a mesh accuracy of 2 units. This configuration allowed for faster simulation run-time while still providing reliable results. We utilized a mesh source for the source setup, selecting the x-axis as the injection axis, which excited the electromagnetic field within the simulation domain. The amplitude was set to 1 unit with a phase of 0. We considered only the fundamental mode for our simulation. The simulation software also allowed us to find the refractive index of the structure using an index monitor, which was essential to consider the effect of the refractive indices of gold and graphene that were added to the structure.

In choosing the analyte, we conducted a study for different milk concentrations[62]. The refractive index varied according to the percentage of water in the solution, and changes in the sensitivity and FOM of the sensor were monitored. Table 1 shows the relationship between different fractions of water and the density of milk, which was responsible for the variation in the refractive index of this study. The refractive index of the analyte varied from 1.347 to 1.352. This variation of the refractive index results in the shift in peaks of transmission spectra, resulting in the sensitivity of the proposed device.

**Table 1.** Refractive indices of analyte used in this study.

| Water (%) | The density of Milk (g/cm$^3$) | RI |
|---|---|---|
| 83 | 1.043455406 | 1.352 |
| 84 | 1.040794925 | 1.351 |
| 85 | 1.038147977 | 1.350 |
| 86 | 1.038147977 | 1.349 |
| 87 | 1.032894267 | 1.348 |
| 88 | 1.030287302 | 1.347 |

*3.2 Design Optimization*

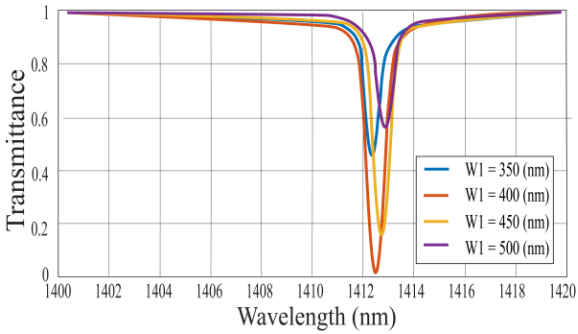
(a) Transmittance

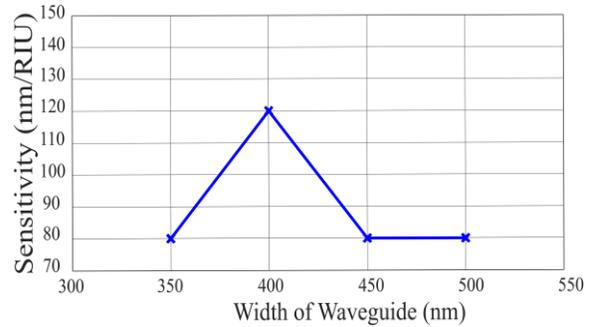
(b) Sensitivity

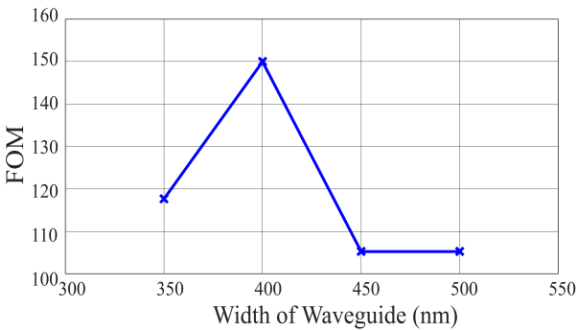
(c) FOM

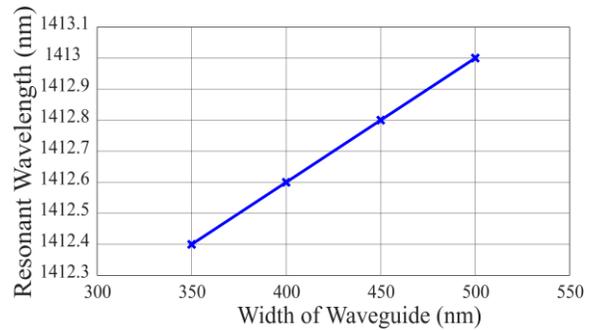
(d) Resonant Wavelength

**Fig. 2.** (a) Transmission characteristics, (b) Sensitivity, (c) FOM, (d) Resonant wavelength shift t with change in width of rectangular waveguide, $W_1$.

For the SOI MRR, the parameters of the unoptimized structure are given in Table 2. Various structural factors influence a system's transmittance, and these parameters are investigated in the following parts. To observe the variation in transmission spectrum and performance parameters, a physical parameter from Table 2 is altered, keeping the rest the same. After each optimization, we retain that value for the subsequent parameter optimization. The analyte's refractive index was set to 1.348 during simulation to find optimized physical parameters.

Initially, the width of the rectangular waveguide, $W_1$, varies from 350 nm to 500 nm. The transmission spectrum is shown in Fig. 2(a) for four distinct values, keeping all other parameters constant, as shown in Table 2. The dip occurs at the resonant wavelengths. When increasing $W_1$, the value of sensitivity and FOM increases uniformly, reaching a maximum at 400 nm and then decreasing to become constant eventually. As a result, the chosen width for our optimized structure was 400 nm. These performance parameters are varied in Fig. 2(b),(c), respectively. Fig. 2(d) shows that the redshift increases with the rectangular waveguide's width. At a width of 400nm, propagation losses are lower in the waveguide. With reduced losses, more light energy can be effectively delivered to the ring resonator, improving sensitivity and FOM.

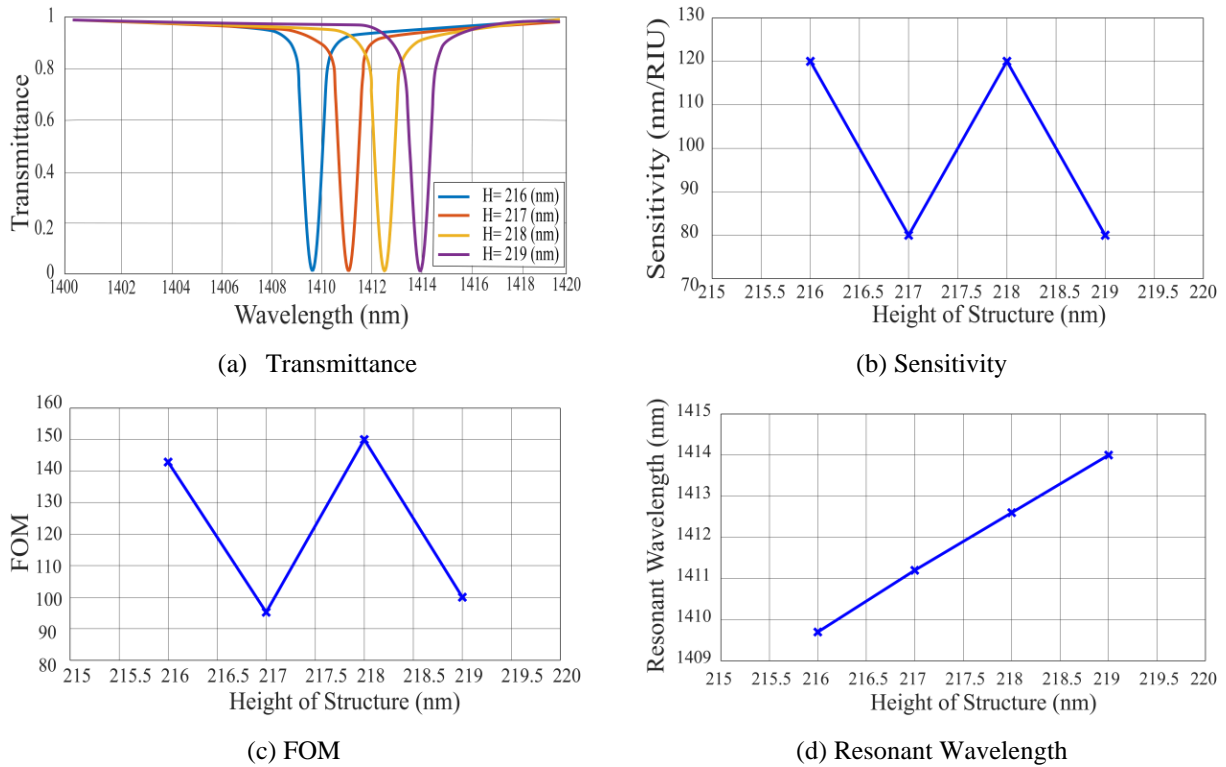

**Fig. 3.** (a) Transmission characteristics, (b) Sensitivity, (c) FOM, (d) Resonant wavelength shift with a change in Height, $H$.

Secondly, $H$ represents the height of the ring and the rectangular waveguide, which varies from 216 nm to 219 nm with a step size of 1 nm. While testing for this parameter, all other parameters were kept unchanged. Fig. 3(a) represents the transmittance on the vertical axis and wavelength on the horizontal. Variation of *H* results in sensitivity and FOM changes shown in Fig. 3(b),(c). A redshift is seen when we increase the height, illustrated in Fig. 3(d). An increase in the height of the MRR results in a longer effective optical path length for light propagation within the ring. The increased height causes a redshift in the resonance wavelength, accumulating more phases by the light over each round-trip. This accumulation necessitates a longer wavelength for constructive interference and the attainment of resonance. When *H* increases from 216 nm to 217 nm, sensitivity and FOM fall linearly. Sensitivity and FOM rise uniformly after 217 nm before falling again after 218 nm. Therefore, a height of 218 nm allowed us to have the maximum possible sensitivity and FOM for our optimized structure. In this instance, the height nearly matches an integer multiple of the operational wavelength. When the optical modes of the waveguide and the MRR are phase-matched at this height, a resonance condition leads to improved coupling and increased sensitivity and FOM. Even a tiny deviation from the resonance condition can decrease coupling efficiency, resulting in lower sensitivity and FOM.

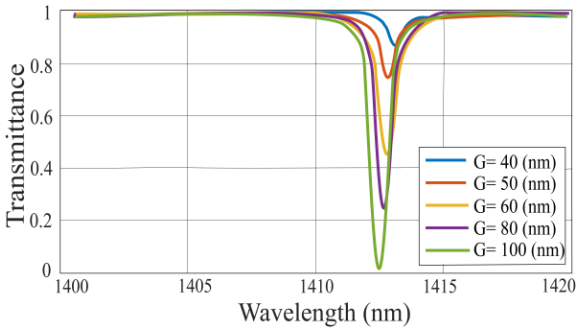

(a) Transmittance

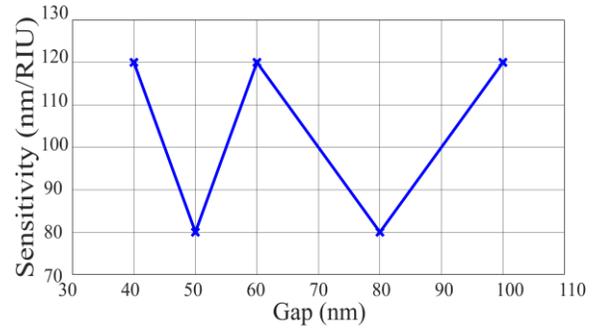

(b) Sensitivity

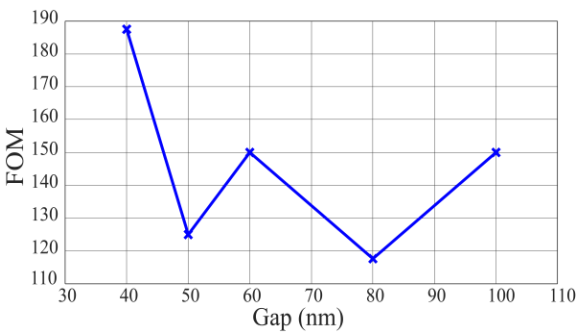

(c) FOM

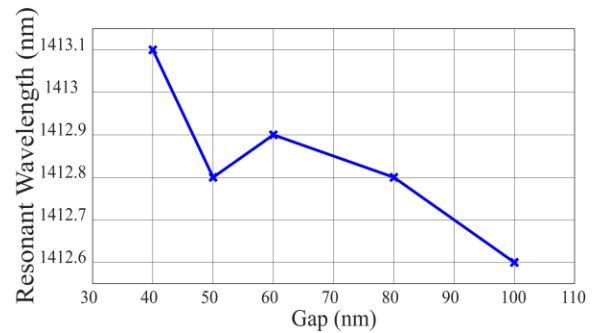

(d) Resonant Wavelength

**Fig. 4.** (a) Transmission characteristics, (b) Sensitivity, (c) FOM, (d) Resonant wavelength shift with a change in coupling gap, *G*.

Thirdly, alongside maintaining consistency across all parameters and varying coupling gap $G$ between the ring and the rectangular waveguide from 40 nm to 100 nm, the transmittance as a function of wavelength is shown in Fig. 4(a). Fig. 4(b), (c) illustrate the changes in sensitivity and FOM, respectively. When the coupling gap increased, it was observed that the performance parameters changed randomly. The sensitivity decreased and increased alternatively, while FOM followed a similar nonlinear trend. The best combination of sensitivity and FOM was observed to be at the gap of 40 nm, which is the coupling gap of our optimized structure. Fig. 4(d) exhibits the shift in resonance peaks. The gap distance affects the phase-matching condition between the waveguide and the MRR. When the phase matching condition is met, there is constructive interference between the light propagating in the waveguide and the resonant modes of the MRR. This constructive interference enhances the coupling efficiency and the sensitivity of the device. With a gap of 40 nm, the phase matching condition is more effectively fulfilled, leading to improved values of performance parameters.

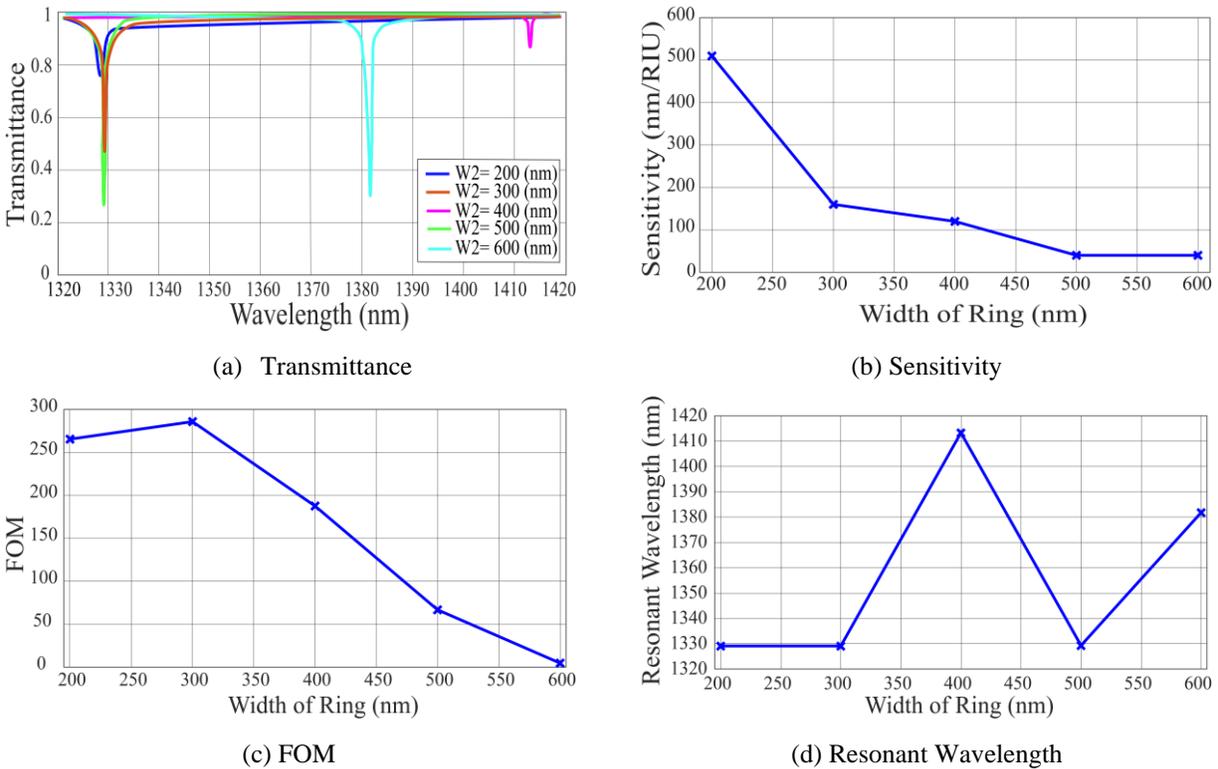

(a) Transmittance  (b) Sensitivity
(c) FOM  (d) Resonant Wavelength

**Fig. 5.** (a) Transmission characteristics, (b) Sensitivity, (c) FOM, (d) Resonant wavelength shift with a change in width of the ring, $W_2$.

Afterward, the width of ring $W_2$ is investigated, and the transmission spectra are obtained from 200 nm to 600 nm for five equal-spaced values; all other parameters were kept unchanged. The transmission is shown in Fig. 5(a), and the associated performance metrics are given in Fig. 5(b),(c). Fig. 5(d) clearly illustrates the resonant peak shifts. A general trend shows that the sensitivity and FOM decrease as $W_2$ increases, although there is a slight increase in FOM from 200 nm to 300 nm. At a ring width of 200nm, the maximum sensitivity and an exceptionally high FOM were observed. Consequently, 200 nm was selected for our optimized ring width. At $W_2 = 200$ nm, the waveguide width is optimized for strong evanescent field overlap and mode confinement. The narrower waveguide confines the mode tightly, resulting in a stronger evanescent field and higher sensitivity to refractive index changes. The more confined mode profile enhances the interaction with the external medium, improving sensitivity and FOM compared to larger waveguide widths.

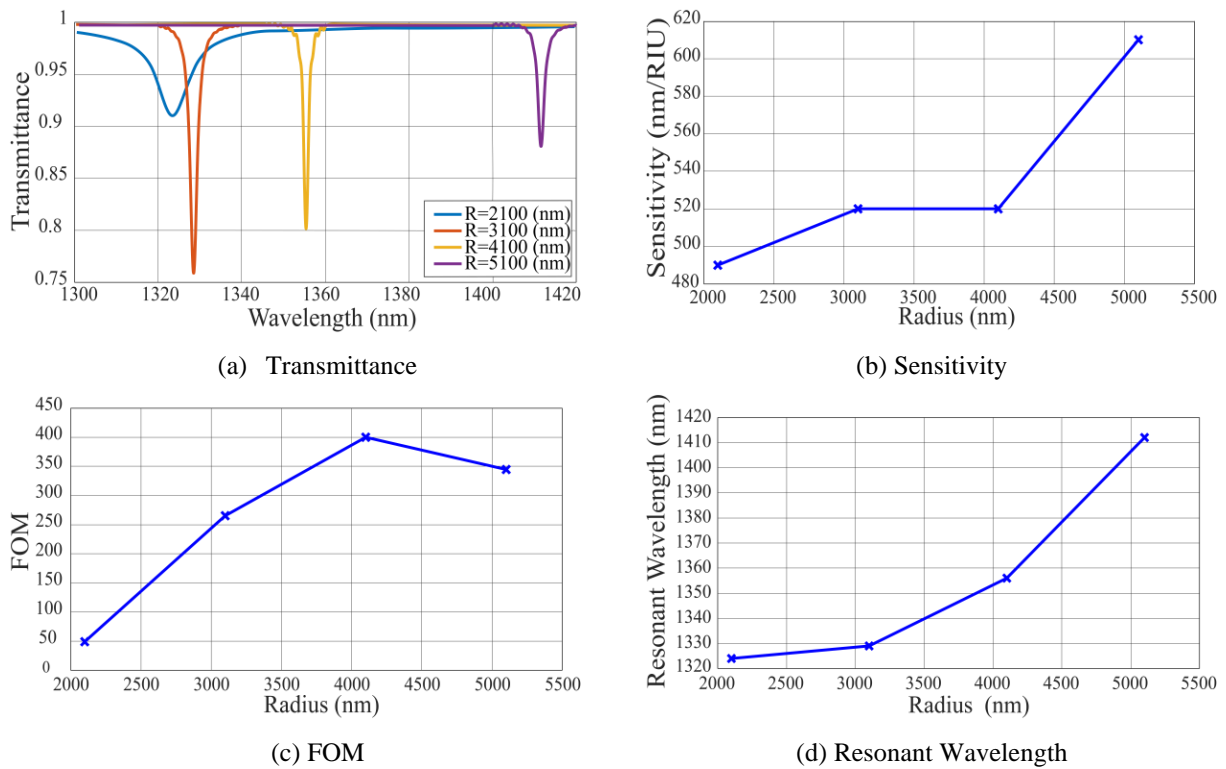

**Fig. 6.** (a) Transmission characteristics, (b) Sensitivity, (c) FOM, (d) Resonant wavelength shift with a change in radius of the ring, *R*.

Lastly, when Fig. 6(a) is observed, the transmission characteristic for variation in the radius of the ring, *R*, can be seen, while all other parameters were kept unaltered. The value is taken from 2100 nm to 5100 nm with an interval of 1000 nm. Fig. 6(b),(c) represent the sensitivity and FOM. As *R* is increased, the value of sensitivity rises almost linearly. However, FOM increases to a radius value of 4100 nm and then decreases. Thus, a ring radius of 5100nm gave the best compromise between sensitivity and FOM. The shift in resonance peak is represented by Fig. 6(d). As *R* increases, the circumference of the ring also increases. This results in a longer path length for the light circulating within the ring resonator. Due to the longer path length, the effective index of the mode within the ring decreases. When the effective index falls, the resonance condition is achieved at longer wavelengths, leading to a redshift in the resonance wavelength according to the relationship in equation (1). For the optimized value of R=5100nm, the mode profile within the MRR can exhibit strong field confinement and evanescent field overlap with the surrounding medium. This enhanced evanescent field overlap enables more efficient interaction between the guided mode in the MRR and the external environment, giving the highest sensitivity.

Therefore, after the optimization, the parameters that gave the best performance resulted in a sensitivity of 610 nm/RIU with a corresponding FOM of 344. Both are highly optimized values obtained from the FDTD analysis of our initial model. The associated parameters are given in Table 2. Our optimized refractive index sensor has outperformed those of recent studies in terms of sensitivity. A good increase in sensitivity is achieved by about 10 percent compared to Kumari et al. [24], which our optimized device outperforms.

**Table 2.** Optimized Geometric Parameters.

| Parameter of Optimized Structure | Representation | Unoptimized Value | Optimized Value | Unit |
|---|---|---|---|---|
| Height | $H$ | 216 | 218 | nm |
| Ring Radius | $R$ | 3100 | 5100 | nm |
| Ring width | $W_2$ | 400 | 200 | nm |
| The gap between the waveguide and ring | $G$ | 100 | 40 | nm |
| Waveguide width | $W_1$ | 400 | 400 | nm |
| The gap between gold nano-disc and micro-ring | $G_g$ | - | 66 | nm |

*3.3 Further enhancement*

At first, a graphene monolayer is added on top of the substrate, further enhancing the SOI MRR's optimized performance. An improvement is observed immediately. The sensitivity increases to 680 nm/RIU, and FOM is reduced to 118. Both are prominent values. We then tested the optimized SOI MRR model with a gold disc inside the MRR to obtain higher sensitivity. A gap, $G_g$, is maintained between the gold disc and the MRR, whose value is set as 66 nm as used by Kumari et al. [24], which gives sensitivity and FOM of 690 nm/RIU and 69, respectively. This results in a hybrid structure, which outperforms all of those in Table 3 and further increases the sensitivity of the optimized model to 730 with a reduction in FOM to 56.15, which is acceptable. Thus, an average increase in sensitivity of 31.53% is obtained. This value is the highest at the time of this study. Fig. 7(a) shows the transmission profile of the optimized structure with the incorporation of gold and graphene at varying refractive indices. Fig. 7(b),(c) indicates the relationship between sensitivity and FOM with varying refractive indices of the analyte in the study. Figure 8(a),(b),(c),(d) shows the graphical representation of the electrical field distribution during resonance for the optimized SOI MRR, with the addition of graphene only to the optimized structure, with the addition of gold only to the optimized system, and finally with the addition of both gold and graphene to the proposed design after optimization. In all the cases, at resonance, the electric field is strongly confined within the ring structure, particularly in the vicinity of the waveguide region. This confinement arises due to the total internal reflection at the waveguide edges, trapping the light within the ring.

The electric field distribution diagram shows high-intensity regions where the electric field is concentrated, typically localized near the waveguide or the coupling region. These regions correspond to the areas where the light is tightly confined and interacts strongly with the surrounding materials, such as the graphene layer or gold nano-discs.

In addition to the work carried out till now, the structure was analyzed after changing the fermi energy level of the graphene layer. The chemical potential of the material was varied from 0.4 eV to 0.7 eV with an increment of 0.05 eV in the simulation software since, in many cases, the fermi energy and the chemical potential are used interchangeably, especially in the context of semiconductors or conductive materials like graphene [63]. The results for sensitivity and FOM remained almost unchanged.

**Table 3.** Comparison of sensitivities with previous publications.

| Paper | Year | Sensitivity (nm/RIU) | FOM (RIU$^{-1}$) |
|---|---|---|---|
| L. Arce et al. [64] | 2010 | 70 | - |
| Nugroho et al. [65] | 2020 | 85.84 | - |
| C. Ciminelli et al. [26] | 2013 | 120 | - |
| M. Grist et al.[52] | 2013 | 142 | - |
| F. Khozeymeh et al. [50] | 2018 | 193 | 3112 |
| T. Claes et al. [66] | 2009 | 298 | - |
| R. Raj et al.[27] | 2018 | 360 | - |

| Y. Wen et al. [67] | 2019 | 500 | 2000 |
| S. Kumari et al. [24] | 2022 | 555 | 154.16 |
| Optimized Basic SOI MRR (This paper) | 2023 | 610 | 344 |
| Proposed Sensor (This paper) | 2023 | 730 | 56.15 |

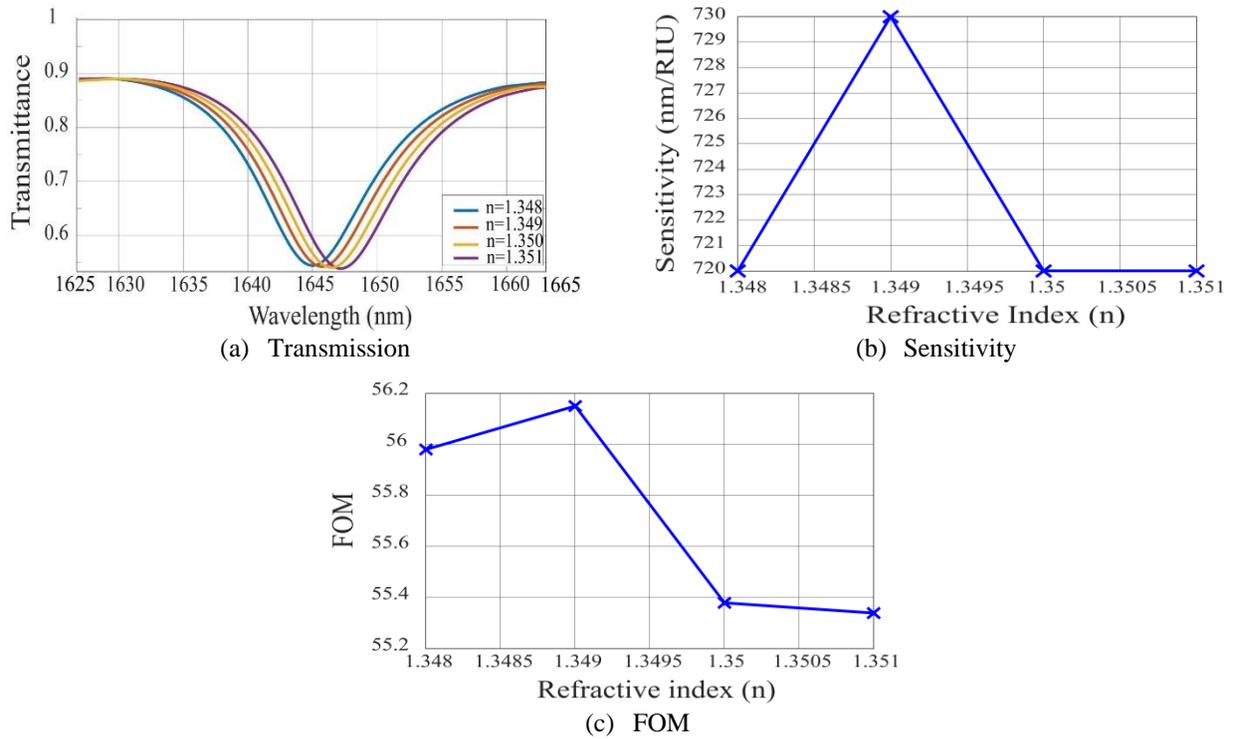

(a) Transmission

(b) Sensitivity

(c) FOM

**Fig. 7.** (a) Transmission profile of the sensor at different refractive indices (b) Sensitivity of the sensor at different refractive indices (c) FOM of the sensor at different refractive indices.

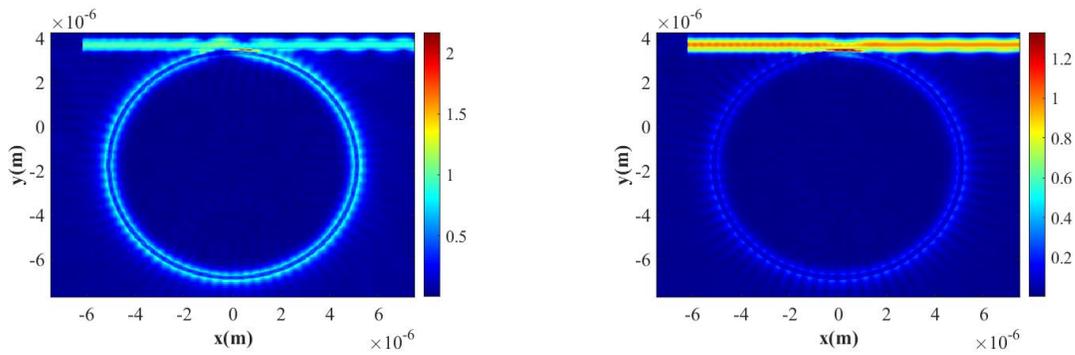

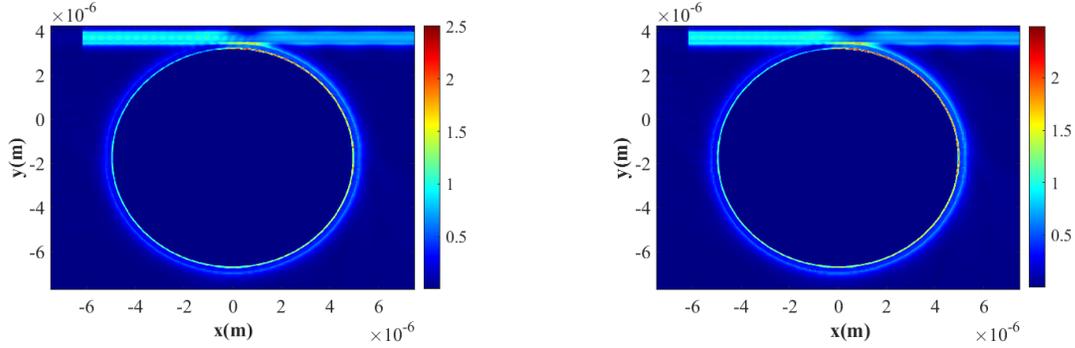
**Fig. 8.** Electric Field profiles for (a) Optimized SIO MRR, (b) Addition of Graphene only to the optimized structure, (c) Addition of Gold only to the optimized structure, (d) Addition of both Graphene and Gold to the optimized structure.

The suggested SOI MRR has improved performance because of the hybrid nature of its design, and it has excellent potential for use in real-world sensing applications. However, this design cannot outperform MIM devices like those stated in articles [37], [38]. The reason for this is enhanced evanescent field interaction in the MIM structure. However, it must be considered that through this study, we have looked to improve the SOI MRR with a better design and fabrication compatibility with existing silicon photonics fabrication processes. The hybrid nature of the proposed sensor is a means to enhance the performance of the existing SOI technology, which makes it quite different from the MIM sensors that may show better performance on paper.

## 4. Suggested method for fabrication:

Fabrication feasibility is extremely important for implementing the sensor in practical applications. Our proposed fabrication method outlines the essential steps that can be taken to manufacture our SOI ring resonator successfully while incorporating both Graphene and Gold.

Graphene must be deposited on a silicon dioxide substrate. The chemical vapor deposition (CVD) technique can accomplish this. This highly feasible method is frequently applied to fabricate graphene on silicon dioxide substrates. In this method, a furnace with two temperature zones is filled with a combination of methane and hydrogen gases. Placed in the low-temperature region, the SiO2 substrate undergoes methane decomposition, resulting in the formation of graphene on the SiO2 surface. The high-temperature region promotes the formation of high-quality graphene growth. Like metal-catalyzed CVD graphene, the process produces continuous, homogeneous graphene films with tiny surface ripples. For producing graphene on SiO2, the two-temperature zone assembly CVD method described is a practical technology that offers benefits, including continuous and uniform film production without the requirement for a problematic transfer step[68].

Following the graphene deposition on the SiO2 substrate, a silicon ring has to be constructed on the graphene monolayer. Physical vapor deposition (PVD) technology at high temperatures can accomplish this. A low-temperature seed layer enhances wetting on deposited CVD graphene and permits uniform silicon growth. However, monolayer graphene patches might be the only place silicon grows, leaving undeveloped multilayer regions. Heterogeneous silicon nucleation on graphene is possible. It may be necessary to perform additional optimization and gain knowledge of the nucleation behavior to grow silicon on graphene in a homogenous state. Although this method is moderately feasible, more knowledge would be needed to streamline the process [69].

After silicon is deposited on graphene using Physical Vapor Deposition (PVD), the silicon can be shaped using lithography and etching. To transfer the desired pattern using lithography, a photoresist is spin-coated onto the substrate, exposed to UV light through a photomask, and then developed to remove the exposed or unexposed areas. Following lithography, undesirable silicon regions are selectively removed while the graphene and photoresist layers are kept intact using etching processes like dry etching (plasma etching) or wet etching. Material compatibility influences the selection of an etching technique and etch selectivity. The newly formed silicon ring is left on the graphene surface after removing the remaining photoresist using a photoresist stripper solution[70].

Lastly, gold nanoparticles (AuNPs) must be deposited to form a disc within the silicon ring. A spontaneous galvanic reaction technique can deposit gold nanoparticles on graphene. This approach is relatively simple and practical for putting gold nanoparticles on graphene. Therefore, this procedure is highly feasible. To achieve this, graphene must be sandwiched between a reductant (such as germanium or copper) and an oxidant (such as HAuCl4). A. During the reduction of AuCl4, graphene serves as a conduit for the flow of electrons. Microscopy methods can be used to see the resultant AuNPs on the graphene surface. Changing the AuCl4- solution's concentration may alter the size and homogeneity of the AuNPs. A quick increase in current during electrochemical experiments supports charge transmission through the graphene basal plane[71].

## 5. Conclusion:

Our suggested design for a highly sensitive SOI ring resonator can be used as a refractive index sensor. We determined the optimal values for sensitivity and FOM by optimizing several vital parameters. Furthermore, adding a graphene monolayer and a gold disc resulted in a hybrid structure that outperformed previous designs, achieving a sensitivity of 730 nm/RIU and an FOM of 56.15, which was the highest obtained considering our optimization method. The novelty in our work comes from implementing a graphene layer and a gold nano-disc together on our optimized SOI MRR. Our results highlight the potential of this device as a high-performance on-chip refractive index nano-sensor. The results of our study, with the incorporation of graphene and gold into the structure of a regular optimized MRR, can prove to be a significant step towards developing more accurate and effective sensors for various uses, including chemical and biological sensing. Future studies on SOI-based RI sensors can be done, building upon our suggested design and analyzing the performance parameters.

## Funding

No funding was used to support this work.

## Data Availability

No data was used for the research described in the article.

## Declaration of competing interest

The authors declare that they have no known competing financial interests or personal relationships that could have appeared to influence the work reported in this paper.

## References

[1] A. Yacoubian, "Optical Sensing," *Opt. Essentials*, pp. 107–123, 2018, doi: 10.1201/9781315216324-10.
[2] E. Kasper and J. Yu, "Microring Resonators," *Silicon-Based Photonics*, vol. 29, no. 24, pp. 49–93, 2020, doi: 10.1201/9781315156514-4.
[3] P. Colinge, "2 Basics of Silicon-on-Insulator (SOl) Technology," 2004.
[4] W. Bogaerts *et al.*, "Fabrication of photonic crystals in silicon-on-insulator using 248-nm deep UV lithography," *IEEE J. Sel. Top. Quantum Electron.*, vol. 8, no. 4, pp. 928–934, 2002, doi: 10.1109/JSTQE.2002.800845.
[5] M. R. Rakhshani and M. Rashki, "Metamaterial perfect absorber using elliptical nanoparticles in a multilayer metasurface structure with polarization independence," *Opt. Express*, vol. 30, no. 7, p. 10387, 2022, doi: 10.1364/oe.454298.
[6] M. R. Rakhshani, "Narrowband Plasmonic Absorber Using Gold Nanoparticle Arrays for Refractive Index

[7] M. R. Rakhshani and M. Rashki, "Numerical Simulations of Metamaterial Absorbers Employing Vanadium Dioxide," *Plasmonics*, vol. 17, no. 3, pp. 1107–1117, 2022, doi: 10.1007/s11468-021-01563-x.

[8] M. R. Rakhshani, "Three-Dimensional Polarization-Insensitive Perfect Absorber Using Nanorods Array for Sensing and Imaging," *IEEE Sens. J.*, vol. 20, no. 23, pp. 14166–14172, 2020, doi: 10.1109/JSEN.2020.3008489.

[9] X. Li *et al.*, "Sensitive label-free and compact biosensor based on concentric silicon-on-insulator microring resonators," *Appl. Opt.*, vol. 48, no. 25, pp. 3–7, 2009, doi: 10.1364/AO.48.000F90.

[10] L. Hajshahvaladi, H. Kaatuzian, and M. Danaie, "A high-sensitivity refractive index biosensor based on Si nanorings coupled to plasmonic nanohole arrays for glucose detection in water solution," *Opt. Commun.*, vol. 502, no. May 2021, p. 127421, 2022, doi: 10.1016/j.optcom.2021.127421.

[11] A. Densmore *et al.*, "A silicon-on-insulator photonic wire based evanescent field sensor," *IEEE Photonics Technol. Lett.*, vol. 18, no. 23, pp. 2520–2522, 2006, doi: 10.1109/LPT.2006.887374.

[12] J. Becker, A. Trügler, A. Jakab, U. Hohenester, and C. Sönnichsen, "The optimal aspect ratio of gold nanorods for plasmonic bio-sensing," *Plasmonics*, vol. 5, no. 2, pp. 161–167, 2010, doi: 10.1007/s11468-010-9130-2.

[13] C. Y. Zhao, P. Y. Chen, P. Y. Li, and C. M. Zhang, "Numerical analysis of effective refractive index bio-sensor based on graphene-embedded slot-based dual-micro-ring resonator," *Int. J. Mod. Phys. B*, vol. 34, no. 17, 2020, doi: 10.1142/S0217979220501453.

[14] L. Qiao, G. Zhang, Z. Wang, G. Fan, and Y. Yan, "Study on the fano resonance of coupling M-type cavity based on surface plasmon polaritons," *Opt. Commun.*, vol. 433, no. June 2018, pp. 144–149, 2019, doi: 10.1016/j.optcom.2018.09.055.

[15] M. R. Rakhshani and M. A. Mansouri-Birjandi, "High sensitivity plasmonic refractive index sensing and its application for human blood group identification," *Sensors Actuators, B Chem.*, vol. 249, pp. 168–176, 2017, doi: 10.1016/j.snb.2017.04.064.

[16] M. R. Rakhshani and M. A. Mansouri-Birjandi, "Engineering Hexagonal Array of Nanoholes for High Sensitivity Biosensor and Application for Human Blood Group Detection," *IEEE Trans. Nanotechnol.*, vol. 17, no. 3, pp. 475–481, 2018, doi: 10.1109/TNANO.2018.2811800.

[17] R. Zafar, S. Nawaz, G. Singh, A. D'Alessandro, and M. Salim, "Plasmonics-Based Refractive Index Sensor for Detection of Hemoglobin Concentration," *IEEE Sens. J.*, vol. 18, no. 11, pp. 4372–4377, 2018, doi: 10.1109/JSEN.2018.2826040.

[18] M. A. Jabin *et al.*, "Surface Plasmon Resonance Based Titanium Coated Biosensor for Cancer Cell Detection," *IEEE Photonics J.*, vol. 11, no. 4, 2019, doi: 10.1109/JPHOT.2019.2924825.

[19] S. P. Apell, G. W. Hanson, and C. Hägglund, "High optical absorption in graphene," 2012, [Online]. Available: http://arxiv.org/abs/1201.3071

[20] X. Zhu *et al.*, "Enhanced light-matter interactions in graphene-covered gold nanovoid arrays," *Nano Lett.*, vol. 13, no. 10, pp. 4690–4696, 2013, doi: 10.1021/nl402120t.

[21] W. L. Barnes, A. Dereux, and T. W. Ebbesen, "Surface plasmon subwavelength optics," *Nature*, vol. 424, no. 6950, pp. 824–830, 2003, doi: 10.1038/nature01937.

[22] R. Al Mahmud, M. O. Faruque, and R. H. Sagor, "A highly sensitive plasmonic refractive index sensor based on triangular resonator," *Opt. Commun.*, vol. 483, p. 126634, 2021, doi: 10.1016/j.optcom.2020.126634.

[23] D. I. Yakubovsky, A. V. Arsenin, Y. V. Stebunov, D. Y. Fedyanin, and V. S. Volkov, "Optical constants and structural properties of thin gold films," *Opt. Express*, vol. 25, no. 21, p. 25574, 2017, doi: 10.1364/oe.25.025574.

[24] S. Kumari and S. M. Tripathi, "Hybrid Plasmonic SOI Ring Resonator for Bulk and Affinity a Bio - sensing Applications," *Silicon*, vol. 14, no. 17, pp. 11577–11586, 2022, doi: 10.1007/s12633-022-01877-3.

[25] D. T. Debu, P. K. Ghosh, D. French, and J. B. Herzog, "Surface plasmon damping effects due to Ti adhesion layer in individual gold nano-discs," *Opt. Mater. Express*, vol. 7, no. 1, p. 73, 2017, doi: 10.1364/ome.7.000073.

[26] C. Ciminelli, F. Dell'Olio, D. Conteduca, C. M. Campanella, and M. N. Armenise, "High performance SOI microring resonator for biochemical sensing," *Opt. Laser Technol.*, vol. 59, pp. 60–67, 2014, doi: 10.1016/j.optlastec.2013.12.011.

[27] R. R. Singh, S. Kumari, A. Gautam, and V. Priye, "Glucose Sensing Using Slot Waveguide-Based SOI Ring Resonator," *IEEE J. Sel. Top. Quantum Electron.*, vol. 25, no. 1, pp. 1–8, 2018, doi: 10.1109/JSTQE.2018.2879022.

[28] Y. Chen *et al.*, "Sensing performance analysis on Fano resonance of metallic double-baffle contained MDM waveguide coupled ring resonator," *Opt. Laser Technol.*, vol. 101, pp. 273–278, 2018, doi:

Sensing," *IEEE Sens. J.*, vol. 22, no. 5, pp. 4043–4050, 2022, doi: 10.1109/JSEN.2022.3142655.

10.1016/j.optlastec.2017.11.022.

[29] W. Zhou *et al.*, "Polarization-independent and omnidirectional nearly perfect absorber with ultra-thin 2D subwavelength metal grating in the visible region," *Opt. Express*, vol. 23, no. 11, p. A413, 2015, doi: 10.1364/oe.23.00a413.

[30] M. W. Chen, Y. F. Chau, and D. P. Tsai, "Three-dimensional analysis of scattering field interactions and surface plasmon resonance in coupled silver nanospheres," *Plasmonics*, vol. 3, no. 4, pp. 157–164, 2008, doi: 10.1007/s11468-008-9069-8.

[31] S. Khani and M. Hayati, "An ultra-high sensitive plasmonic refractive index sensor using an elliptical resonator and MIM waveguide," *Superlattices Microstruct.*, vol. 156, no. March, p. 106970, 2021, doi: 10.1016/j.spmi.2021.106970.

[32] S. Khani and M. Hayati, "Optical sensing in single-mode filters base on surface plasmon H-shaped cavities," *Opt. Commun.*, vol. 505, no. September 2021, p. 127534, 2022, doi: 10.1016/j.optcom.2021.127534.

[33] F. Moradiani, A. Farmani, M. H. Mozaffari, M. Seifouri, and K. Abedi, "Systematic engineering of a nanostructure plasmonic sensing platform for ultrasensitive biomaterial detection," *Opt. Commun.*, vol. 474, no. February, p. 126178, 2020, doi: 10.1016/j.optcom.2020.126178.

[34] L. Hajshahvaladi, H. Kaatuzian, M. Moghaddasi, and M. Danaie, "Hybridization of surface plasmons and photonic crystal resonators for high-sensitivity and high-resolution sensing applications," *Sci. Rep.*, vol. 12, no. 1, pp. 1–15, 2022, doi: 10.1038/s41598-022-25980-y.

[35] F. Hu, F. Chen, H. Zhang, L. Sun, and C. Yu, "Sensor based on multiple Fano resonances in MIM waveguide resonator system with silver nanorod-defect," *Optik (Stuttg).*, vol. 229, no. December 2020, p. 166237, 2021, doi: 10.1016/j.ijleo.2020.166237.

[36] F. Chen and W. X. Yang, "Pressure sensor based on multiple Fano resonance in metal–insulator–metal waveguide coupled resonator structure," *J. Opt. Soc. Am. B*, vol. 39, no. 7, p. 1716, 2022, doi: 10.1364/josab.461472.

[37] R. Zafar and M. Salim, "Enhanced Figure of Merit in Fano Resonance-Based Plasmonic Refractive Index Sensor," *IEEE Sens. J.*, vol. 15, no. 11, pp. 6313–6317, 2015, doi: 10.1109/JSEN.2015.2455534.

[38] Y. Sharma, R. Zafar, and S. K. Metya, "Split Ring Resonators-Based Plasmonics," *IEEE Sens. J.*, vol. 21, no. 5, pp. 6050–6055, 2021.

[39] F. Chen, "Nanosensing and slow light application based on Fano resonance in waveguide coupled equilateral triangle resonator system," *Optik (Stuttg).*, vol. 171, pp. 58–64, 2018, doi: 10.1016/j.ijleo.2018.03.135.

[40] L. Hajshahvaladi, H. Kaatuzian, and M. Danaie, "A very high-resolution refractive index sensor based on hybrid topology of photonic crystal cavity and plasmonic nested split-ring resonator," *Photonics Nanostructures - Fundam. Appl.*, vol. 51, no. May, p. 101042, 2022, doi: 10.1016/j.photonics.2022.101042.

[41] M. R. Islam, A. N. M. Iftekher, I. Marshad, N. F. Rity, and R. U. Ahmad, "Analysis of a dual peak dual plasmonic layered LSPR-PCF sensor – Double peak shift sensitivity approach," *Optik (Stuttg).*, vol. 280, no. February, 2023, doi: 10.1016/j.ijleo.2023.170793.

[42] A. Samusenko *et al.*, "A SiON microring resonator-based platform for biosensing at 850 nm," *J. Light. Technol.*, vol. 34, no. 3, pp. 969–977, 2016, doi: 10.1109/JLT.2016.2516758.

[43] Z. Tu, D. Gao, M. Zhang, and D. Zhang, "High-sensitivity complex refractive index sensing based on Fano resonance in the subwavelength grating waveguide micro-ring resonator," *Opt. Express*, vol. 25, no. 17, p. 20911, 2017, doi: 10.1364/oe.25.020911.

[44] P. S. Science *et al.*, "ce pte d M us," *Mater. Des.*, vol. 11, no. 20, pp. 5035–5040, 2021, [Online]. Available: https://doi.org/10.1016/j.matdes.2020.109338

[45] M. Farrokhfar, S. Jarchi, and A. Keshtkar, "Multilayer metamaterial graphene sensor with high sensitivity and independent on the incident angle," *Optik (Stuttg).*, vol. 265, no. June, p. 169536, 2022, doi: 10.1016/j.ijleo.2022.169536.

[46] Z. Xiong *et al.*, "Terahertz Sensor with Resonance Enhancement Based on Square Split-Ring Resonators," *IEEE Access*, vol. 9, pp. 59211–59221, 2021, doi: 10.1109/ACCESS.2021.3073043.

[47] B. Xiao, J. Zhu, and L. Xiao, "Tunable plasmon-induced transparency in graphene metamaterials with ring–semiring pair coupling structures," *Appl. Opt.*, vol. 59, no. 20, p. 6041, 2020, doi: 10.1364/ao.394942.

[48] F. A. Said, P. S. Menon, S. Shaari, and B. Y. Majlis, "FDTD Analysis on Geometrical Parameters of Bimetallic Localized Surface Plasmon Resonance-Based Sensor," *Proc. - Int. Conf. Intell. Syst. Model. Simulation, ISMS*, vol. 2015-Octob, pp. 242–245, 2015, doi: 10.1109/ISMS.2015.12.

[49] Y. F. Xiao, B. Min, X. Jiang, C. H. Dong, and L. Yang, "Coupling whispering-gallery-mode microcavities with modal coupling mechanism," *IEEE J. Quantum Electron.*, vol. 44, no. 11, pp. 1065–1070, 2008, doi: 10.1109/JQE.2008.2002088.

[50] F. Khozeymeh and M. Razaghi, "Characteristics optimization in single and dual coupled silicon-on-insulator ring (disc) photonic biosensors," *Sensors Actuators, B Chem.*, vol. 281, no. October 2018, pp. 998–1008, 2019, doi: 10.1016/j.snb.2018.11.017.

[51] M. Iqbal *et al.*, "Label-Free Biosensor Arrays Based on Silicon," *IEEE J. Sel. Top. Quantum Electron.*, vol. 16, no. 3, pp. 654–661, 2010.

[52] S. M. Grist *et al.*, "Silicon photonic micro-disc resonators for label-free biosensing," *Opt. Express*, vol. 21, no. 7, p. 7994, 2013, doi: 10.1364/oe.21.007994.

[53] A. Bozzola, S. Perotto, and F. De Angelis, "Hybrid plasmonic-photonic whispering gallery mode resonators for sensing: A critical review," *Analyst*, vol. 142, no. 6, pp. 883–898, 2017, doi: 10.1039/c6an02693a.

[54] R. Arefin, O. Faruque, R. Al Mahmud, and R. H. Sagor, "Design of a tunable ring resonator with enhanced quality factor," *Proc. 9th Int. Conf. Electr. Comput. Eng. ICECE 2016*, no. 1, pp. 369–372, 2017, doi: 10.1109/ICECE.2016.7853933.

[55] J. Zhou, W. Wang, Y. Wang, J. Feng, and J. Guo, "A highly-sensitive NaCl concentration sensor based on a compact silicon-on-insulator micro-ring resonator," *AOPC 2015 Opt. Optoelectron. Sens. Imaging Technol.*, vol. 9674, p. 96742F, 2015, doi: 10.1117/12.2201034.

[56] Z. Fang *et al.*, "Plasmon-induced doping of graphene," *ACS Nano*, vol. 6, no. 11, pp. 10222–10228, 2012, doi: 10.1021/nn304028b.

[57] D. T. Nurrohman and N. F. Chiu, "A review of graphene-based surface plasmon resonance and surface-enhanced raman scattering biosensors: Current status and future prospects," *Nanomaterials*, vol. 11, no. 1, pp. 1–30, 2021, doi: 10.3390/nano11010216.

[58] Z. Wei *et al.*, "Active plasmonic band-stop filters based on graphene metamaterial at THz wavelengths," *Opt. Express*, vol. 24, no. 13, p. 14344, 2016, doi: 10.1364/oe.24.014344.

[59] F. Chen, D. Yao, and Y. Liu, "Graphene-metal hybrid plasmonic switch," *Appl. Phys. Express*, vol. 7, no. 8, 2014, doi: 10.7567/APEX.7.082202.

[60] F. Chen, H. Zhang, L. Sun, and C. Yu, "Tunable plasmonic properties of graphene ribbon for hypersensitive nanosensing," *Optik (Stuttg).*, vol. 196, no. July, p. 163139, 2019, doi: 10.1016/j.ijleo.2019.163139.

[61] P. B. Johnson and R. W. Christy, "Optical Constant of the Nobel Metals," *Phys. Rev. B*, vol. 6, no. 12, pp. 4370–4379, 1972.

[62] U. Biswas, J. K. Rakshit, and G. K. Bharti, "Design of photonic crystal microring resonator based all-optical refractive-index sensor for analyzing different milk constituents," *Opt. Quantum Electron.*, vol. 52, no. 1, 2020, doi: 10.1007/s11082-019-2140-1.

[63] B. Wei and S. Jian, "A Nanoscale Fano Resonator by Graphene-Gold Dipolar Interference," *Plasmonics*, vol. 13, no. 6, pp. 1889–1895, 2018, doi: 10.1007/s11468-018-0703-9.

[64] C. Lerma Arce, K. De Vos, T. Claes, K. Komorowska, D. Van Thourhout, and P. Bienstman, "Silicon-on-insulator microring resonator sensor integrated on an optical fiber facet," *IEEE Photonics Technol. Lett.*, vol. 23, no. 13, pp. 890–892, 2011, doi: 10.1109/LPT.2011.2143704.

[65] H. S. Nugroho *et al.*, "Silicon on insulator-based microring resonator and Au nanofilm Krestchmann-based surface plasmon resonance glucose sensors for lab-on-a-chip applications," *Int. J. Nanotechnol.*, vol. 17, no. 1, pp. 29–41, 2020, doi: 10.1504/IJNT.2020.109348.

[66] T. Claes, J. G. Molera, K. De Vos, E. Schacht, R. Baets, and P. Bienstman, "Label-free biosensing with a slot-waveguide-based ring resonator in silicon on insulator," *IEEE Photonics J.*, vol. 1, no. 3, pp. 197–204, 2009, doi: 10.1109/JPHOT.2009.2031596.

[67] Y. Wen *et al.*, "High sensitivity and FOM refractive index sensing based on Fano resonance in all-grating racetrack resonators," *Opt. Commun.*, vol. 446, no. April, pp. 141–146, 2019, doi: 10.1016/j.optcom.2019.04.068.

[68] S. C. Xu *et al.*, "Direct synthesis of graphene on SiO2 substrates by chemical vapor deposition," *CrystEngComm*, vol. 15, no. 10, pp. 1840–1844, 2013, doi: 10.1039/c3ce27029g.

[69] G. Lupina, J. Kitzmann, M. Lukosius, J. Dabrowski, A. Wolff, and W. Mehr, "Deposition of thin silicon layers on transferred large area graphene," *Appl. Phys. Lett.*, vol. 103, no. 26, 2013, doi: 10.1063/1.4858235.

[70] Y. Huang, G. T. Paloczi, A. Yariv, C. Zhang, and L. R. Dalton, "Fabrication and replication of polymer integrated optical devices using electron-beam lithography and soft lithography," *J. Phys. Chem. B*, vol. 108, no. 25, pp. 8606–8613, 2004, doi: 10.1021/jp049724d.

[71] Y. Park, J. Y. Koo, S. Kim, and H. C. Choi, "Spontaneous Formation of Gold Nanoparticles on Graphene by Galvanic Reaction through Graphene," *ACS Omega*, vol. 4, no. 19, pp. 18423–18427, 2019, doi: 10.1021/acsomega.9b02691.